\documentclass[preprint,showpacs,superscriptaddress,amsmath,amssymb]{revtex4}

\begin{document}

\preprint{}
\title{Relations for Four Nucleons in a Single j Shell -- \\ Using 9j
Symbols
to Get Rid of 9j Symbols}
\author{L. Zamick}

\affiliation{Department of Physics and Astronomy, Rutgers University,
New Brunswick, New Jersey 08903, USA }

\date{\today}
\pacs{21.600Fw}

\begin{abstract}
In previous works we considered the number of pairs of particles
 with angular momentum J$_{12}$ for systems with total angular momentum
zero.
Here
 we extend the the work to states of any total angular momentum.  We find
 simplified relations satisfied  for a system of four nucleons with isospins
 zero and two.
\end{abstract}

\maketitle

\section{INTRODUCTION}
We previously considered the problem of the number of pairs of nucleons with
angular momentum  J$_{12}$  for systems with total angular momentum I= zero
e.g. ground states of even-even Ti isotopes. [1,2,3]  We could get general
expressions  for  any J$_{12}$ pairs and any total angular momentum I, but
we
obtained very simplified expressions for the cases where I was equal to zero
and J$_{12}$ was also zero  i.e. the number of J$_{12}$ = zero pairs in the
I=zero
ground states of even-even Ti isotopes.  For $^{44}$Ti we went further and
obtained simplified expressions for I=0 and all even J$_{12}$.

In this work we show that we can get some simplified expressions for all I
in many but not all cases. The cases where we succeed include even J$_{12}$
pairs for states of total isospin T=0 and T=2.

\section{METHOD}
As in the work of Zamick, Lee and Mekjian [3] we start out by noting that
  that for the j$^{2}$ configuration, states of even angular momentum have
   isospin T=1 while states of odd angular momentum have isospin T=0.
 We then consider an interaction which is a constant for all odd J and a
 different constant for all even J.

\begin{equation}
V = a(1-(-1)^{J})/2 + b(1 + (-1)^{J})/2
\end{equation}

 This same interaction can be written using isospin variables.

\begin{equation}
V = a(\frac{1}{4} - t(1) \cdot t(2)) + b(\frac{3}{4} + t(1) \cdot t(2))
\end{equation}

The general expression for the matrix element of any interaction
  for $^{44}$Ti in the single j shell is

\newpage
\begin{eqnarray}
& &<[(j^{2})^{J_{p}} (j^{2})^{J_{N}}]^{I} V[(j^{2})^{J^{\prime}_{P}}
(j^{2})^{J^{\prime}_{N}}]^{I}> \nonumber \\
& = & (E(J_{P}) + E(J_{N})) \delta_{J_{P}^{\prime}}
\delta_{J_{N},J^{\prime}_{N}} \nonumber \\
& + & 4 \sum_{J_{A}J_{B}} <(jj)^{J_{P}} (jj)^{J_{N}} \mid (jj)^{J_{A}}
(jj)^{J_{B}} >^{I}\nonumber \\
& & < (jj)^{J^{\prime}_{P}}(jj)^{J^{\prime}_{N}} \mid (jj)^{J_{A}}
(jj)^{J_{B}} >
^{I} E(J_{B})
\end{eqnarray}

   Where

\begin{equation}
E(J_{B}) = <(j^{2})^{J_{B}} V(j^{2})^{J_{B}}>
\end{equation}

The last term in Eq. (3) has a product of two unitary 9j symbols.

Let us now consider the case where b=0 i.e. the T=0 interaction
is ``a'' and the T=1 interaction is zero.  Let us take a to be one.

    The result for the matrix element of eq. (3) is now

\begin{eqnarray}
& & <[(j^{2})^{J_{P}} (j^{2})^{J_{N}}]^{I} V[(j^{2})^{J^{\prime}_{P}}
(j^{2})^{J^{\prime}_{N}}]^{I}> \nonumber \\
& = & 2 \delta_{J_{P}J^{\prime}_{P}} \delta_{J_{N}J^{\prime}_{N}} \nonumber
\\
& + & 2 <(jj)^{J_{P}} (jj)^{J_{N}} \mid (jj)^{J^{\prime}_{P}}
(jj)^{J^{\prime}_{N}}>^{I}
\end{eqnarray}

  We can also evaluate the eigenvalues of this ``Hamiltonian'' using isospin
   considerations.

  For A nucleons we have

\begin{equation}
\sum_{i<j} (\frac{1}{4} - t(i) \cdot t(j)) = \frac{A^{2}}{8} + \frac{A}{4} -
\frac{1}{2} T (T+1)
\end{equation}

   For A=4 the eigenvalues are 3, 2, and zero for isospins T=0, 1 and 2
   respectively.

     A wave function of $^{44}$Ti can,in the single j shell, be represented
      as a column vector, one of the entries being D$^{I\alpha}(Jp,Jn)$
  which is the probability amplitude that in a state of total angular
  momentum I, the protons couple to Jp and the neutrons to Jn.

   We have the orthogonality and completeness conditions

\begin{equation}
\sum_{J_{P}J_{N}} D^{I\alpha}(J_{P}J_{N}) D^{I^{\prime}\alpha^{\prime}}
(J_{P}J_{N}) = \delta_{II^{\prime}} \delta_{\alpha \alpha^{\prime}}
\end{equation}

\begin{equation}
\sum_{\alpha} D^{I\alpha}(J_{P}J_{N}) D^{I\alpha} (J^{\prime}_{P}J^{\prime}
_{N}) = \delta_{J_{P} J^{\prime}_{P}} \delta_{J_{N}J^{\prime}_{N}}
\end{equation}

  From Eq. (5) we now have the following ``eigenvalue equation'' for a state
of total angular momentum I and total isospin T

\begin{equation}
\sum_{J^{\prime}_{P}J^{\prime}_{N}}[2 \delta_{J_{P}J^{\prime}_{P}}
\delta_{J_{N}J^{\prime}_{N}} + 2 <(jj)^{J_{P}} (jj)^{J_{N}} \mid
(jj)^{J^{\prime}_{P}} (jj)^{J^{\prime}_{N}}>^{I}] D^{I}(J^{\prime}_{P}
J^{\prime}_{N}) = \lambda^{T} D^{I}(J_{P}J_{N})
\end{equation}

Alternately

\begin{equation}
2 \sum_{J_{P}J_{N}} <(jj)^{J_{P}}(jj)^{J_{N}} \mid (jj)^{J^{\prime}_{P}}
(jj)^{J^{\prime}_{N}}>^{I} D^{I}(J^{\prime}_{P}J^{\prime}_{N}) =
(\delta_{T,0} + 0 \delta_{T,1} -2 \delta_{T,2}) D^{I}(J_{P}J_{N})
\end{equation}

\section{NUMBER of PROTON-NEUTRON  PAIRS WITH ANGULAR MOMENTUM J$_{12}$}

To get the number of neutron-proton pairs with angular momentum J$_{12}$
  we refer to Eq (3) where the matrix element of the nuclear Hamiltonian
  is written as a sum of a neutron-neutron part, a proton-proton part and
  a proton-neutron part. Clearly the last term is of interest here.

     To get the number of p-n pairs of angular momentum J$_{12}$ we set
all E(J$_{b}$)
 to zero except for E(J$_{12}$), which is set equal to one. We then get the
 expectation value of this Hamiltonian with the wave function of a state
  of interest.

We get

\begin{equation}
\# \ of \ p-n \ pairs \ (J_{12}) = \sum_{J_{A}} \mid 2 \sum_{J_{P}J_{N}}
<(j^{2})^{J_{P}}(j^{2})^{J_{N}} \mid (j^{2})^{J_{A}}(j^{2})^{J_{12}}>
D^{I}(J_{P}J_{N}) \mid ^{2}
\end{equation}

  Inside the square of Eq (10) we have something that looks like the left
hand
  of Eq (9b).  But we have to be careful. The sum in Eq (10) is over even
  Jp, Jn -- indeed D(Jp,Jn) is defined for only even values of the angular
   momenta in Eq. (10) however Ja and J$_{12}$ can be either even or odd.

    We  will now show that in certain cases, despite the above, we can
simplify
     the expressions for the number of p-n pairs.  We can only do this for
    even J$_{12}$.  But still we have to worry about the fact that Ja can be
    odd.  We must consider three different isospin cases T=0, 1 and 2.

   For T=0 and 2 the following is true.

\begin{equation}
D^{I}(J_{P}J_{N}) = (-1)^{I} D^{I} (J_{N}J_{P})
\end{equation}

This is a consequence of charge symmetry.

On the other hand for T=1 we have

\begin{equation}
D^{I}(J_{P}J_{N}) = (-1)^{(I+1)} D^{I}(J_{N}J_{P})
\end{equation}

Eq 12 insures that terms with Ja odd will cancel out in the sum over
   Jp,Jn.  We can therefore use Eq (9b) for these two cases.  We find
 \# of p-n pairs (J$_{12}$ even) T=0 or 2

\begin{equation}
= \sum_{J_{A}} \mid D^{I}(J_{A}J_{12} \mid^{2} (\delta_{T,0} +4
\delta_{T,2})
\end{equation}

We cannot get a correspondingly simple expression for states with isospin
T=1.

By summing Eq. (14) over even $J_{12}$ we find that for T=0 states the total
number of even $J_{12}$ pairs is one, and since the total number of $J_{12}$
pairs is four, the total number of odd $J_{12}$ pairs is three.

A more general result was previously obtained using isospin considerations,
see
A(12) and A(13) in ref. [2].  There it is shown that the total number of
T$_{12}$=0 (and hence odd $J_{12}$) pairs is equal to

\begin{equation}
\frac{A^{2}}{8} + \frac{A}{4} - \frac{T(T+1)}{2}
\end{equation}

The total number of T$_{12}$=1 (and hence even $J_{12}$ pairs), including nn
and pp as well as np is equal to

\begin{equation}
\frac{3A^{2}}{8} - \frac{3A}{4} + \frac{T(T+1)}{2}
\end{equation}

In the above A is the number of valence nucleons.  The above results hold
for
all total angular momenta I and for all Ti isotopes, in the single j
approximation.

In the single j approximation we get the following results for A=4 for the
number of pairs (nn + pp + np)
\bigskip

(\# of odd pairs, \# of even pairs), T=0 (3, 3), T=1 (2, 4), T=2 (0, 6).

For T=2 there are no odd J$_{12}$ pairs.  This can be understood by the fact
that the T=2 state is the double analog of a state in $^{44}$Ca where we are
dealing with identical particles i.e. neutrons.  Two neutrons in a single j
shell must have isospin one and even angular momentum.

The above results are especially simple in the 1s shell.  Two neutrons and
two protons close this shell so the total isospin must be zero.  The even
pairs have angular momentum J$_{12}$ = 0, while the odd ones have
J$_{12}$ = 1.  Thus there are three J$_{12}$ = 0 pairs and three
J$_{12}$ = 1 pairs in this case.

\end{document}